\documentclass[10pt,twocolumn]{IEEEtran}
\usepackage[utf8]{inputenc}
\usepackage[english]{babel}
\usepackage{graphicx}%
\graphicspath{{./figures/}}
\usepackage{multirow}%
\usepackage{amsmath,amssymb,amsfonts}%
\usepackage{amsthm}%
\usepackage{mathrsfs}%

\usepackage{xcolor}%
\usepackage{textcomp}%
\usepackage{manyfoot}%
\usepackage{booktabs}%
\usepackage{listings}%
\usepackage{tikz,lipsum,lmodern}
\usepackage{tikz-3dplot}
\usetikzlibrary{positioning,arrows,calc,math,angles,quotes}
\usepackage{blochsphere}
\usepackage{listofitems} 
\usetikzlibrary{shapes,shapes.geometric,backgrounds,fit}
\usepackage{amsmath,amssymb,amsthm}

\theoremstyle{definition}

\usepackage{enumerate}
\usepackage{qcircuit}

\usepackage{caption}
\usepackage{subcaption}
\usepackage[most]{tcolorbox}

\tcolorboxenvironment{myitemize}{blanker,
before skip=6pt,after skip=6pt,
borderline west={3mm}{0pt}{white}}

\usepackage{todonotes}

\newcommand{\ket}[1]{\ensuremath{\left|#1\right\rangle}} 
\newcommand{\bra}[1]{\ensuremath{\left\langle#1\right|}} 

\renewcommand{\bf}[1]{\ensuremath{\mathbf{#1}}}

\begin{document}

\title{A unifying primary framework for quantum graph neural networks from quantum graph states}


\author{Ammar~Daskin
\IEEEauthorblockN{
                      \thanks{adaskin25@gmail.com}  
}
}


\maketitle

\begin{abstract}
Graph states are used to represent mathematical graphs as quantum states on quantum computers. They can be formulated through stabilizer codes or directly quantum gates and quantum states.
In this paper we show that a quantum graph neural network model can be understood and realized based on graph states.
We show that they can be used either as a parameterized quantum circuits to represent neural networks or as an underlying structure to construct graph neural networks on quantum computers.

\end{abstract}
\section{Introduction}

\subsection{Graph Neural Networks \cite{liu2022introduction,zhou2020graph,wu2020comprehensive,sanchez-lengeling2021a,bronstein2021geometric}}
Artificial neural networks are described through neurons (nodes) connected to each other through a multi layered structure.  In neural networks each node is connected to every node in the subsequent layers. This can be depicted as follows: \\
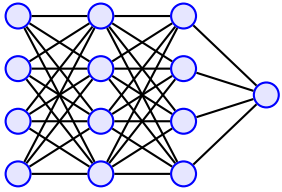
\begin{center}
    
\tikzstyle{mynode}=[thick,draw=blue,fill=blue!10,circle,minimum size=3]
\begin{tikzpicture}[x=2.2cm,y=1.4cm, scale=0.5]
  \readlist\Nnod{4,4,4,1} 
  \foreachitem \N \in \Nnod{ 
    \foreach \i [evaluate={\x=\Ncnt; \y=\N/2-\i+0.5; \prev=int(\Ncnt-1);}] in {1,...,\N}{ 
      \node[mynode] (N\Ncnt-\i) at (\x,\y) {};
      \ifnum\Ncnt>1 
        \foreach \j in {1,...,\Nnod[\prev]}{ 
          \draw[thick] (N\prev-\j) -- (N\Ncnt-\i); 
        }
      \fi 
    }
  }  
\end{tikzpicture}
\end{center}

On the other hand, graph neural networks \cite{scarselli2008graph} (see Ref.\cite{sanchez-lengeling2021a} for an introductory tutorial or Ref.\cite{velivckovic2023everything} for a very short overview.) describe neural network structure through specified graphs where the connection between subsequent layers are defined by the connection properties of the graph. 
They are also used as building blocks to solve combinatorial optimization problems \cite{cappart2023combinatorial}.
In general, graphs represent some real world problems by equating vertices to the features and relationships between features to the edges. 
Therefore, for a feature matrix $X=[x_1, \dots, x_{|V|}]$, each vertex represents a feature in the data set.
To build an example graph neural network, we consider the following  graph:
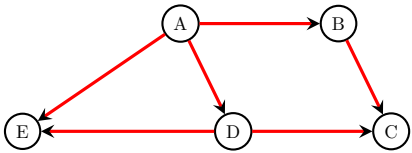
\begin{center}
    \tikzset{
pics/layer/.style args={#1/#2}{
code={
    \begin{scope}[
    vnode/.style={circle,thick,draw, fill=#1!30},
    >={stealth[black]},
                  enode/.style={fill=#1!5,circle, midway},
                  every edge/.style={draw=red,very thick}]
        \node[vnode] (A) at (0,0) {A};
        \node[vnode]  (B) at (0,3) {B};
        \node[vnode]  (C) at (2.5,4) {C};
        \node[vnode]  (D) at (2.5,1) {D};
        \node[vnode]  (E) at (2.5,-3) {E};
        \path [->] (A) edge node {} (B)
        (B) edge node {} (C)
        (A) edge node {} (D)
        (D) edge node {} (C)
        (A) edge node {} (E)
        (D) edge node {} (E);
    \end{scope}
    }}}
\tdplotsetmaincoords{-35}{90}
\begin{tikzpicture}[tdplot_main_coords,thick,scale=0.7, every node/.style={transform shape}]
\node (layer1) at  (0,0,0){};
\pic[]  at (layer1){layer={white/}};
\end{tikzpicture}

\end{center}

\begin{figure}[ht]
    \centering
    \tikzset{
pics/layer/.style args={#1/#2}{
code={
    \begin{scope}[
    vnode/.style={circle,thick,draw, fill=#1!30},
    >={stealth[black]},
                  enode/.style={fill=#1!5,circle, midway},
                  every edge/.style={draw=red,very thick}]
        \node[vnode] (A) at (0,0) {A};
        \node[vnode]  (B) at (0,3) {B};
        \node[vnode]  (C) at (2.5,4) {C};
        \node[vnode]  (D) at (2.5,1) {D};
        \node[vnode]  (E) at (2.5,-3) {E};
        \path [->] (A) edge node {} (B)
        (B) edge node {} (C)
        (A) edge node {} (D)
        (D) edge node {} (C)
        (A) edge node {} (E)
        (D) edge node {} (E);
    \end{scope}
    \scoped[on background layer]
    \node [tape, tape bend height=1cm, fill=#1!30, minimum width=5cm,
    fit=(A)(B)(C)(D)(E), rotate=-20, inner sep=5pt,opacity=0.4] {#2};
    }}}
\tdplotsetmaincoords{-35}{90}
\begin{tikzpicture}[tdplot_main_coords,thick,scale=0.9, every node/.style={transform shape}]
\node (layer3) at  (8,0,-1){};
\node (layer2) at  (4,0,-0.5){};
\node (layer1) at  (0,0,0){};
\pic[]  at (layer3){layer={cyan/Layer 3}};
\pic[]  at (layer2){layer={green/Layer 2}};
\pic[]  at (layer1){layer={blue/Layer 1}};
          \begin{scope}[on background layer,
    >={stealth[black]},
                  enode/.style={fill=black!5,circle, midway},
                  edgeblue/.style={draw=red!90,very thick,dashed},
                  edgered/.style={draw=red!90,very thick,dashed},
                  edgegreen/.style={draw=red!90,very thick,dash dot}
                  ]
        \path [->,edgeblue] (A) edge node {} ($(B)+(layer2)$)
        (A) edge node {} ($(D)+(layer2)$)
        (A) edge node {} ($(E)+(layer2)$);
        \path [->,edgegreen] ($(D)+(layer2)$) edge node {} ($(E)+(layer3)$)
        ($(D)+(layer2)$) edge node {} ($(C)+(layer3)$);
        \path [->,edgegreen] ($(B)+(layer2)$) edge node {} ($(C)+(layer3)$);
    \end{scope}
\end{tikzpicture}
    \caption{An example graph neural network: Each node accumulates information through the nodes in the subsequent layers where only nodes in the neighborhood of the node in the original graph are considered. As an example, the vertex A is connected to E, D, and B in the original graph. Therefore, it is updated using these vertices in the previous layer.  }
    \label{fig:gnn}
\end{figure}
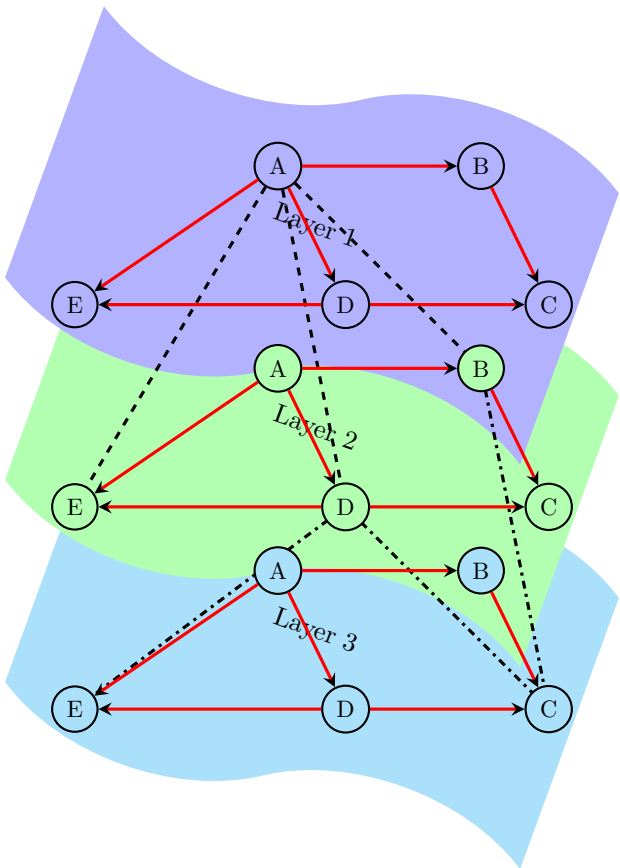

A GNN with this graph can be built by layering this graph as in the neural networks and using the connectivity of the original graph to adjust information flow (the connection) between layers.
An example with three such layers is shown in Fig.\ref{fig:gnn} where the information flow between layers is represented for the node $A$ in the first layer. The information of $A$ accumulates through the nodes $E,B, D$ in the second layer since there is an edge starting from $A$ and ending with any of these nodes in the original graph.
Based on the newly obtained information, the state of a node or an edge can be updated via a trainable model.   

This construction can be considered as a generalization of the neural networks since we can also describe neural networks by using the above description. Therefore, we can say that graph neural networks are at least as powerful as the neural networks in solving different kinds of problems \cite{xu2018powerful}. 
On the other hand, since the connectivity in a problem can be represented by different isomorphic graphs, there is no guarantee that all isomorphic graphs would produce the same deep neural networks. 
This means the graph neural networks are not permutation invariant \cite{sanchez-lengeling2021a}. Since there is no certain way to choose the original graph and hence to determine the optimum construct, one can say that the training task becomes more difficult. 

\subsubsection{Pooling and message passing (update) operations}
The operations in GNNs can be summarized as \emph{pooling} and \emph{message passing} \cite{gilmer2017neural} where either the information is flowed from the edges to the nodes or from the nodes to the edges: 
In message passing, the information in nodes are updated by using a differential function applied to the messages sent by their immediate neighbors. 
This layer mainly changes the graph into a new representation. That is why it is also called the permutation equivariant layer in the literature. Mathematically, in message passing layer of GNNs, the  message passing for a node $u$ with the neighborhood $N(u)$ can be described as a function of the node and edge features by the followings \cite{bronstein2021geometric}:
\begin{equation}
\label{eq:classical_msgpassing}
h_u = \phi \left( \bf{x}_u, \bigoplus \psi (\bf{x}_u, \bf{x}_v, \bf{e}_{uv})\right),
\end{equation}
where $v\in N(u)$, $\bf{x}_u, \bf{x}_v,$ and $\bf{e}_{uv}$ are the features for nodes $u$ and $v$ and the edge $(u,v)$.
In addition, $\phi$ and $\psi$ are the differentiable functions commonly used in neural networks such as the ReLU function. These functions are also called update and message passing functions. And $\bigoplus$ represents a permutation invariant aggregation operator such as sum, max, or mean.  
Similarly, there are also the attention  and the convolution based update functions which are incorporated with the function definitions of $\phi$ and $\psi$ (see Ref.\cite{velivckovic2023everything}).

The pooling layer could be local or global. It basically applies pooling operation (the neighborhood aggregation) first to the local neighborhoods and then to the resulting graph globally to obtain a readout value which is used to compare learning outcome.
There are different pooling operations such as top-$k$ pooling\cite{gao2019graph} or self attention pooling \cite{lee2019self}. 

Here, as it can be observed from the above definitions, the construction of a neural network though a graph gives more freedom in choosing and describing a neural network model. Therefore, one can apply different pooling or message passing strategies based on the problem.

\subsection{Different Level of Tasks}
In summary, a problem can be solved by a GNN in general by the following steps \cite{sanchez-lengeling2021a}: 
\begin{itemize}
    \item We take a graph as an input to the model. In some GNN models, it takes input from neural networks. 
    \item Apply GNN layers. This can be a single graph convolution network\cite{wu2019simplifying,zhang2019graph} or a multi-layered graph neural network.
    \item Output a transformed graph
    \item Then, apply a classification or another network layer to out a prediction.
\end{itemize}
The details of these steps are adjusted or changed based on the considered graph tasks. In literature, 
these tasks are categorized as follows\cite{sanchez-lengeling2021a,velivckovic2023everything}:
\subsubsection{Graph-Level Tasks}
The graph level tasks can be supervised classification or unsupervised clustering like problems. For instance, it may be related to identifying a given graph based on whether it includes some structures or associating one of the given labels to the graph. The example applications include chemical property prediction such as message passing graph networks applied to quantum chemistry problems \cite{gilmer2017neural} and molecular fingerprints that classify molecules for pharmaceuticals\cite{duvenaud2015convolutional}.
\subsubsection{Node-Level Tasks}
These tasks involve predicting target labels for each node: e.g. disease prediction \cite{sun2020disease} or problems related to social networks \cite{fan2019graph}.
\subsubsection{Edge-Level Tasks}
Similarly to the node level task, in this case we deal with the edges: Either we predict the property of a link or its existence (see Ref.\cite{zhang2018link,li2024evaluating} for the definition of the link prediction problem and its solution with GNNs).
\subsection{Quantum Graph Neural Network Models}

In recent years, quantum research in graph neural networks is as vibrant as classical research. 
There are a few quantum graph neural networks models proposed and applied to various problems: This includes Ref.\cite{verdon2019quantum} which has described a parameterized circuit with sequence of parameterized quantum Hamiltonians as an equivalent to the classical graph neural network. Using the same remedy, it has also formulated quantum versions of many different classical graph neural network models such as graph recurrent, graph convolution, and spectral graph convolutional networks.  
Ref.\cite{hu2022design} has proposed a graph convolution network  based on variational quantum circuits where each link between nodes is established by using a controlled rotation $Y$ gate. 
Ref.\cite{innan2024financial} has applied the same proposed design in \cite{hu2022design} to the financial fraud detection problem. 
In addition,
variational approaches are used for the simulation of particle interactions \cite{collis2023physics} and material search \cite{ryu2023quantum}, a hybrid approach is used for particle track reconstruction\cite{tuysuz2021hybrid}, . 
Ref.\cite{ai2022decompositional} has described an ancillary based model where graph initializing, convolution, and other layers are parameterized with the controlled rotation gates $X, Y,$ and $Z$. 
This model is tested on various real world datasets and has overall performed well in those tests against the chosen classical models.
There are also models such as graph convolutional networks \cite{zhang2019quantum,zheng2021quantum,chen2021hybrid} and a quantum feature embedding described for classical graph neural networks\cite{xu2023quantum}.

\subsection{Motivation and Contribution}

The problems solved with graphs are in general difficult which requires processing graphs with enormous sizes. Therefore, graph sizes in real world applications tend to be inherently large and continue to increase every year \cite{sanchez-lengeling2021a}.

As mentioned above, there are already a few quantum graph neural network models. 
These models can be summarized as parameterized quantum circuits where some quantum gates are determined from the graphs. Although this encompasses the employment of a given graph into the neural network model, it is immediately not clear how it can be generalized for different graphs and problems using different formulations. In addition, it is not clearly stated how these models are different from the general quantum parameterized circuits used as neural network models.

Quantum graph states, as explained in the following section, are considered the quantum equivalent of mathematical graphs. Although many proposed quantum graph neural networks have some fruits of graph states in their definitions, to the best of our knowledge, they are not directly formulated through the graph states. 
Therefore, we believe it is important to have a formalism of graph neural networks based on the graph states. 
This makes their implementation general and their design intuition more understandable.

\section{Quantum Graph Neural Network Model Based on Graph States}
\subsection{Quantum Graph States}
Graph states \cite{schlingemann2001stabilizer,schlingemann2001quantum}   are instances of stabilizer codes \cite{gottesman1997stabilizer} which are used to design error correcting codes. 
They can be used to study entanglement in mathematical graphs and formulate quantum solutions to the problems represented by these graphs(see Ref.\cite{hein2006entanglement}). 
Given a graph $G=(V, E)$ with the sets of vertices $V =\{v_1, \dots, v_n\}$ and edges $E \subseteq \{ (u, v) | (u, v) \in V^2$ and $u\neq v\} $:
\begin{equation}
    \ket{G} = \prod _{(u, v)  \in E} Uz(u,v) \ket{+}^{\otimes n},
\end{equation}
where $Uz(u,v)$ is a $Z$ gate that is controlled by the qubit $u$ and applies $Z$ to the qubit $v$. This basically establishes an entanglement between qubits $u$ and $v$ when there is an edge $(u, v)$ in graph $G$. 
One can also specify the amount of the entanglement by using a general phase gate instead of a $Z$ gate:
\begin{equation}
\label{eq:uzuvw}
    Uz(u, v, w) = e^{-iw\sigma_z^u\sigma_z^v},
\end{equation}
where $\sigma_z$ is one of the Pauli matrices $\sigma_x$, $\sigma_y$, and $\sigma_z$ and $w$ corresponds to the weight on the edge $(u, v)$ of the graph. Here note that, the resulting weighted graph states \ket{G} can be decomposed concisely into the standard basis with $N$ number of $W_i$ vectors: 
\begin{equation}
\begin{split}
    \ket{G} = &\frac{1}{\sqrt{N}} \sum_{W\subseteq V} \prod_{u, v\in E} Uz(u, v, w_{uv}) \ket{W}_z  \\ 
        =   & \frac{1}{\sqrt{N}} \sum_{W\subseteq V} e^{i\frac{1}{2} W\cdot A \cdot W} \ket{W}_z,
\end{split}
\end{equation}
 where $A$ is the adjacency matrix for the graph, $W$ is the matrix of the basis set, and $\ket{W}_z = \ket{W_1}\dots \ket{W_N}$.

This entanglement for a given set of pairs of qubits can be also established with the stabilizer codes. For each vertex $v$, we define a stabilizer operator with the Pauli matrices $\sigma_{x,y,z}$:
\begin{equation}
    S_v = \sigma_x^{(v)} \prod_{u\in N(v)} \sigma_z^{(u)},
\end{equation}
where $N(v)$ is the neighborhood of $v$. The graph state $\ket{G}$ is the common eigenvector to the set of independent commuting operators $\mathcal{S} = \{S_{v_1}, \dots, S_{v_n}\}$. That means for any $S_{v} \in \mathcal{S}$, 
\begin{equation}
    S_v\ket{G} = \ket{G}.
\end{equation}
When the state \ket{G} is measured according to the pattern $S_v$ on $x-$direction and $z-$direction with the outcomes $m_x^v=\pm 1$ and $m_z^u=\pm 1$, $S_v$ puts the following constraint to the outcome:
\begin{equation}
    m_x^v\prod m_z^u = 1.
\end{equation}
Here, note that the generalization to weighted graphs is not possible in the scope of Stabilizer formalism within the Pauli group, however, the amount of the entanglement can still be related to weights in the graphs (See Ref.\cite{hein2006entanglement} for more properties.).

In this paper, we will use the graph state \ket{G(W)} which represents a graph with the weights given by the matrix $W$. 

\subsection{Incorporating Data in \ket{G}}
Quantum machine learning models are based on variational (parameterized) quantum circuits. 
These models either feed data into the quantum gates or uses a quantum state vector \ket{\bf{x}} representing normalized data vector $\bf{x}$ or uses a single qubit 
and its state \ket{x_i} to represent the $i^\text{th}$ feature in vector $\bf{x}$.
Note that one can also prepare the data matrix as a density operator by using $\sum_i \ket{\bf{x_i}}\bra{\bf{x}}$.

Since we describe GNNs through quantum graph states, the gates $U(u,v,w)$ describes the operation for the edge ($u$,$v$) with the weight $w$ and each qubit state \ket{+} describes a node of the graph.
To incorporate data into these states, we will use the above-mentioned formalism where \ket{x_i} describes the qubit state for the $i^\text{th}$ feature. 
In this case, instead of \ket{+}, each qubit state would be the following:
\begin{equation}
    \ket{x_i} = \alpha_i \ket{0} +\beta_i \ket{1}.
\end{equation}

As a result, the graph state \ket{G} reads as:
\begin{equation}
     \ket{G} = \prod _{(u, v)  \in E} Uz(u,v,w_{uv})  \bigotimes_i^n \ket{x_i}.
\end{equation}
Note that the dimension of \ket{x_i} may be different based on the considered data. In that case one can use multiple qubits for each feature (or vertex in the graph).

Also note that we can further rewrite this by using a parameterized single quantum gate $R_y(\theta_i)$ on each qubit $i$ with an angle value $\theta_i$:
\begin{equation}
    \ket{G} = \prod _{(u, v)  \in E} Uz(u,v,w_{uv})  \bigotimes_i^n R_y(\theta_i) \ket{0}.
\end{equation}
In this formalism, by using the above equation, one can try to classically optimize parameters $\theta_i$s and the weights $w_{uv}$ as done in the standard quantum machine learning models based on parameterized quantum circuits.

\subsection{Representing Layers}
The layers in classical GNNs can be considered as the number of updates applied to the graph before getting the final outcome.
In quantum version, we have multiple choices to design equivalent formulation: 
\begin{enumerate}
    \item One can represent this whole process as a quantum state with a fixed number of layers.
\begin{equation}
\label{eq:gnn1}
   \ket{GNN_1} = \frac{1}{\sqrt{m}}\sum_i^m \ket{G_i},
\end{equation}
where \ket{G_i} represents the $i^\text{th}$ layer in the defined $m-$layer GNN.
Then by applying some operations this whole \ket{GNN} can be turned into a desired neural network which represents the trained model.
\item We can also use the following:
\begin{equation}
\label{eq:gnn2}
   \ket{GNN_2} =  \ket{G_1}\dots \ket{G_m}. 
\end{equation}
In this case, the trained model can be obtained by controlled operations between subsequent registers: i.e., \ket{G_i} is updated by using $G_{i-1}$ and so on. Then, \ket{G_m} becomes the final solution.

\item One can also start with a graph \ket{G_1} and  change this into a desired graph \ket{G_m} by applying quantum operations adjusted and chosen based on some measurement results:
\begin{equation}
    \label{eq:gnn3}
    \ket{GNN_3} =  \ket{G_1} \rightarrow \ket{G_2} \rightarrow \dots \rightarrow \ket{G_m}.
\end{equation}
In this case \ket{G_m} represents the final trained neural network.
\end{enumerate}

Although the above definitions seem different, one can show that they are equivalent in the formalism of quantum information which allows measurements along with computations. However, the number of computations and their computational complexities are different since they require different numbers of qubits and operations.

In the following subsections, we are going to define the pooling and message passing  operations used in classical GNNs. 
Here note that we will describe these operations in general terms since their specifications can be adjusted based on the considered graph problem which can be either an edge, a node, or a global graph level task.

\subsection{Quantum Pooling}
\subsubsection{Pooling with Measurements and the State Collapsing}In \ket{GNN}, a qubit represents the state of a vertex in the graph. 
Therefore, the pooling (a type of accumulation) operation can be implemented by using various measurements on the qubits.
If we have $n$-qubit state, the qubits first can be put into groups based on their neighborhoods. 
Then starting from measuring the local neighborhood, successive measurements at the end yield a readout vector of dimension determined by the final number of qubits (if we are trying to predict a single global binary value, then the last qubit output determines the readout.).  

Here, the important point is that if the local neighborhood requires measuring some $\tilde{n}$ number of qubits, 
it requires $O(2^{\tilde{n}})$ complexity to obtain a tomography of the
$\tilde{n}-$qubit state with accuracy  $O(2^{-\tilde{n}})$. 
Therefore,  the size of the local neighborhoods for pooling should be kept as small as possible with this pooling method.

The second important point is that classical pooling methods can be also based on nonlinear functions such as the sigmoid function. 
Since \ket{G} is designed in Hadamard bases with the phases determined by the weights. 
The difference in the measurement outcome of \ket{0} and \ket{1} on any qubit is determined by the cosine of the phases. 
This provides a periodic activation function (e.g. see Ref.\cite{daskin2018simple}) in contrast to the aperiodic functions like the sigmoid or step functions. 
By feeding the local measurement results into the chosen aperiodic function, one can classically compute the nonlinear aperiodic functions from the same measurement results.

\subsubsection{Pooling with the phase estimation}
Instead of measurement one can use an ancilla register to compute the phases in the local neighborhoods.
In this case, each $Uz$ operation is controlled by the qubit representing $Uz$'s local group in the ancilla register. 
A neural network model is also described based on this idea \cite{daskin2018simple}.

\subsubsection{Pooling with Controlled Rotation or NOT Gates}
A controlled rotation operation on any qubit involves their amplitudes $\alpha_i$, $\beta_i$s, and their phases.
If it is not necessary to output a single value from the whole graph, then we can describe pooling as a rotation operation.
Then, any measurement on the qubit involves all the amplitudes in a summation which can be considered as the final step of the pooling operation.
\subsection{Quantum Message Passing}
As in the classical case given in Eq.\eqref{eq:classical_msgpassing}, the quantum message passing can be designed through quantum operations. However, since the constructed graph \ket{G} represents an entangled quantum state any operation on any qubit may potentially affect the other qubits.

Since the whole quantum state \ket{G} involves not only \ket{x_i}s but also $w_{uv}$s, an operation only on the $z-$axis potentially changes only the initial weights. If we apply controlled-$Z$ gates, then this operation involves the amplitudes $\alpha_i$s and $\beta_i$s in \ket{x_i}s, and changes the weights, i.e. the connection (entanglement) between qubits.
Using this intuition, we can describe the following operation as quantum message passing or update operation:
\begin{equation}
    Uz(u,N(u), w_{uN(u)}), 
\end{equation}
where $N(u)$ is the local neighborhood of $u$ and $w_{uN(u)}$ can be -1 for a simple $Z$ gate or a different phase which may be related to the problem type.
Note that this controlled $Z-$ gate can be applied also to the determined local graph invariants. 
\subsection{Formalism for Solving Different Graph Tasks}
\subsubsection{Node Level Tasks}
We prepare each qubit in \ket{x_i} state with the amplitudes $\alpha_i$ and $\beta_i$.
These amplitudes change during the graph evolution with the operations such as pooling and message passing.
Therefore, any measurement on the $y-$ axis gives us the qubit state \ket{0} or \ket{1} which can be used for binary classification.
For non-binary cases, one needs to use more than one qubit: For four node classes, we use two qubits for each node.

\subsubsection{Edge Level Tasks}
Since the edge weights are created by using the controlled rotation $Z$ gates ($U_z$ in \eqref{eq:uzuvw}) dimension, the information of the updated weights are related to the phases. 
This information can be obtained by measuring qubits on the $z-$ axis: i.e. using a measurement operator as a product of Pauli Z gates.

In addition, using an ancilla register in the Hadamard bases, and controlling related $U_z$ operations by this ancilla, we can obtain the updated phases (direct phase estimation) which can be used for link based predictions.

\subsubsection{Graph Level Tasks}
Graph states describe a graph. 
Therefore, the similarity between quantum states can be obtained through swap-tests.
If we have multiple classes and if we need to determine the class of a given graph state (i.e. global classification of the graphs), we can use a superposition of the classes, and obtain a superposition of the swap tests. 
The measurement result yields the class of the given graph with a probability which can be high if the euclidean separation of the classes are wide enough. 

\subsection{Computational Complexity}
We have presented three different formalism in Eq.\eqref{eq:gnn1}, \eqref{eq:gnn2}, and \eqref{eq:gnn3} for representing GNN layers as \ket{GNN_1}, \ket{GNN_2}, and \ket{GNN_3}, respectively.
In terms of complexity, the formalism \ket{GNN_3} is the simplest since we have a single network, for the representation, it requires less number of qubits than \ket{GNN_2}. And since it is not a superposition state, obtaining node probabilities or edge values from this state are easier than obtaining from \ket{GNN_1}.

Operations such as pooling and message passing involves two or more qubits and as discussed before, they can be implemented through controlled $Z$ and $Y$ gates.
However, in \ket{GNN_1}, if these operations are global, then they can be applied to all layers in superposition.
One can also simplify these operations in \ket{GNN_2}.

\ket{GNN_3} can be considered a classical equivalent where we process graphs from one layer to another layer in a sequential manner. 
Therefore, in terms of complexity, it cannot provide any significant advantage over the classical models. However, the entanglement between qubits can perfectly mimic the connections in real world data. 
Therefore, this may improve the accuracy of the models.

\subsection{Other Operations on Graphs}
Similar to convolutions in neural networks, there are polynomial filters and Chebnet \cite{defferrard2016convolutional} which are also used building blocks for the graph neural networks \cite{daigavane2021understanding}.
The polynomial filters are described through the graph Laplacian $L$:
\begin{equation}
    L = D-A,
\end{equation}
where $A$ is the adjacency matrix and $D$ is the diagonal degree matrix.
Polynomials of the Laplacian operator is a general $d$-degree polynomial of the matrix $L$::
\begin{equation}
  p_\bf{w}(L) = w_0I+w_1L^1+w_2L^2+\dots w_dL^d = \sum_{i=0}^{d} w_iL^i,  
\end{equation}
where $\bf{w}$ is a vector of coefficients $w_0, \dots, w_d$. If we know the eigendecomposition of $L$ or if $L$ is a unitary matrix, we can generate the above polynomial easily on quantum circuits.
 by using a unitary gate \cite{daskin2022quantum}:
\begin{equation}
\label{eq:UL}
    U_L = \left(\begin{matrix}
        I & 0\\
        0 & L
    \end{matrix}\right).
\end{equation}

\begin{figure}[t]
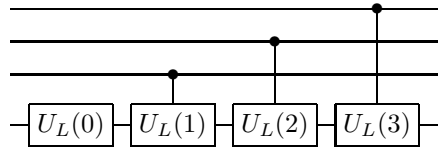

\include{fig_circuitL}
	\caption{ The circuit for construction of the polynomial of the Laplacian operator. $U_L(i)$ represents the $i^\text{th}$ power of the matrix given in Eq.\eqref{eq:UL}}.
 \label{fig:circuitL}
\end{figure}
If we put this gate on a circuit as depicted for four qubits in Fig.\ref{fig:circuitL}, we can obtain the following matrix representation:
\begin{equation}
P(A) = 
\left(\begin{matrix}
I&&&&\\
&L^1&&\\
&&L^2&\\
&&&\ddots&\\
&&&&L^{n-1}
\end{matrix}\right).
\end{equation}

We can include the $w$ vector as in the initial state of this matrix. Therefore, we can implement a quantum filter. 
This operation can be directly applied to the models \ket{GNN} we have described before.

\section{Conclusion}
In this paper, we go through all the necessary steps to define quantum graph neural networks through quantum graph states.
We show how graph neural networks can be represented by using different formulations and discuss how pooling and message passing operations can be applied in these formulations. 
Since graph states are well understood within the quantum information community, they can play a unifying framework and be building blocks to study quantum graph neural networks and understand their power. 

\bibliographystyle{ieeetr}
\bibliography{main}

\begin{thebibliography}{10}

\bibitem{liu2022introduction}
Z.~Liu and J.~Zhou, {\em Introduction to graph neural networks}.
\newblock Springer Nature, 2022.

\bibitem{zhou2020graph}
J.~Zhou, G.~Cui, S.~Hu, Z.~Zhang, C.~Yang, Z.~Liu, L.~Wang, C.~Li, and M.~Sun, ``Graph neural networks: A review of methods and applications,'' {\em AI open}, vol.~1, pp.~57--81, 2020.

\bibitem{wu2020comprehensive}
Z.~Wu, S.~Pan, F.~Chen, G.~Long, C.~Zhang, and S.~Y. Philip, ``A comprehensive survey on graph neural networks,'' {\em IEEE transactions on neural networks and learning systems}, vol.~32, no.~1, pp.~4--24, 2020.

\bibitem{sanchez-lengeling2021a}
B.~Sanchez-Lengeling, E.~Reif, A.~Pearce, and A.~B. Wiltschko, ``A gentle introduction to graph neural networks,'' {\em Distill}, 2021.
\newblock https://distill.pub/2021/gnn-intro.

\bibitem{bronstein2021geometric}
M.~M. Bronstein, J.~Bruna, T.~Cohen, and P.~Veli{\v{c}}kovi{\'c}, ``Geometric deep learning: Grids, groups, graphs, geodesics, and gauges,'' {\em arXiv preprint arXiv:2104.13478}, 2021.

\bibitem{scarselli2008graph}
F.~Scarselli, M.~Gori, A.~C. Tsoi, M.~Hagenbuchner, and G.~Monfardini, ``The graph neural network model,'' {\em IEEE transactions on neural networks}, vol.~20, no.~1, pp.~61--80, 2008.

\bibitem{velivckovic2023everything}
P.~Veli{\v{c}}kovi{\'c}, ``Everything is connected: Graph neural networks,'' {\em Current Opinion in Structural Biology}, vol.~79, p.~102538, 2023.

\bibitem{cappart2023combinatorial}
Q.~Cappart, D.~Ch{\'e}telat, E.~B. Khalil, A.~Lodi, C.~Morris, and P.~Veli{\v{c}}kovi{\'c}, ``Combinatorial optimization and reasoning with graph neural networks,'' {\em Journal of Machine Learning Research}, vol.~24, no.~130, pp.~1--61, 2023.

\bibitem{xu2018powerful}
K.~Xu, W.~Hu, J.~Leskovec, and S.~Jegelka, ``How powerful are graph neural networks?,'' {\em arXiv preprint arXiv:1810.00826}, 2018.

\bibitem{gilmer2017neural}
J.~Gilmer, S.~S. Schoenholz, P.~F. Riley, O.~Vinyals, and G.~E. Dahl, ``Neural message passing for quantum chemistry,'' in {\em International conference on machine learning}, pp.~1263--1272, PMLR, 2017.

\bibitem{gao2019graph}
H.~Gao and S.~Ji, ``Graph u-nets,'' in {\em international conference on machine learning}, pp.~2083--2092, PMLR, 2019.

\bibitem{lee2019self}
J.~Lee, I.~Lee, and J.~Kang, ``Self-attention graph pooling,'' in {\em International conference on machine learning}, pp.~3734--3743, PMLR, 2019.

\bibitem{wu2019simplifying}
F.~Wu, A.~Souza, T.~Zhang, C.~Fifty, T.~Yu, and K.~Weinberger, ``Simplifying graph convolutional networks,'' in {\em International conference on machine learning}, pp.~6861--6871, PMLR, 2019.

\bibitem{zhang2019graph}
S.~Zhang, H.~Tong, J.~Xu, and R.~Maciejewski, ``Graph convolutional networks: a comprehensive review,'' {\em Computational Social Networks}, vol.~6, no.~1, pp.~1--23, 2019.

\bibitem{duvenaud2015convolutional}
D.~K. Duvenaud, D.~Maclaurin, J.~Iparraguirre, R.~Bombarell, T.~Hirzel, A.~Aspuru-Guzik, and R.~P. Adams, ``Convolutional networks on graphs for learning molecular fingerprints,'' {\em Advances in neural information processing systems}, vol.~28, 2015.

\bibitem{sun2020disease}
Z.~Sun, H.~Yin, H.~Chen, T.~Chen, L.~Cui, and F.~Yang, ``Disease prediction via graph neural networks,'' {\em IEEE Journal of Biomedical and Health Informatics}, vol.~25, no.~3, pp.~818--826, 2020.

\bibitem{fan2019graph}
W.~Fan, Y.~Ma, Q.~Li, Y.~He, E.~Zhao, J.~Tang, and D.~Yin, ``Graph neural networks for social recommendation,'' in {\em The world wide web conference}, pp.~417--426, 2019.

\bibitem{zhang2018link}
M.~Zhang and Y.~Chen, ``Link prediction based on graph neural networks,'' {\em Advances in neural information processing systems}, vol.~31, 2018.

\bibitem{li2024evaluating}
J.~Li, H.~Shomer, H.~Mao, S.~Zeng, Y.~Ma, N.~Shah, J.~Tang, and D.~Yin, ``Evaluating graph neural networks for link prediction: Current pitfalls and new benchmarking,'' {\em Advances in Neural Information Processing Systems}, vol.~36, 2024.

\bibitem{verdon2019quantum}
G.~Verdon, T.~McCourt, E.~Luzhnica, V.~Singh, S.~Leichenauer, and J.~Hidary, ``Quantum graph neural networks,'' {\em arXiv preprint arXiv:1909.12264}, 2019.

\bibitem{hu2022design}
Z.~Hu, J.~Li, Z.~Pan, S.~Zhou, L.~Yang, C.~Ding, O.~Khan, T.~Geng, and W.~Jiang, ``On the design of quantum graph convolutional neural network in the nisq-era and beyond,'' in {\em 2022 IEEE 40th International Conference on Computer Design (ICCD)}, pp.~290--297, IEEE, 2022.

\bibitem{innan2024financial}
N.~Innan, A.~Sawaika, A.~Dhor, S.~Dutta, S.~Thota, H.~Gokal, N.~Patel, M.~A.-Z. Khan, I.~Theodonis, and M.~Bennai, ``Financial fraud detection using quantum graph neural networks,'' {\em Quantum Machine Intelligence}, vol.~6, no.~1, pp.~1--18, 2024.

\bibitem{collis2023physics}
B.~Collis, S.~Patel, D.~Koch, M.~Cutugno, L.~Wessing, and P.~M. Alsing, ``{Physics simulation via quantum graph neural network},'' {\em AVS Quantum Science}, vol.~5, p.~023801, 04 2023.

\bibitem{ryu2023quantum}
J.-Y. Ryu, E.~Elala, and J.-K.~K. Rhee, ``Quantum graph neural network models for materials search,'' {\em Materials}, vol.~16, no.~12, p.~4300, 2023.

\bibitem{tuysuz2021hybrid}
C.~T{\"u}ys{\"u}z, C.~Rieger, K.~Novotny, B.~Demirk{\"o}z, D.~Dobos, K.~Potamianos, S.~Vallecorsa, J.-R. Vlimant, and R.~Forster, ``Hybrid quantum classical graph neural networks for particle track reconstruction,'' {\em Quantum Machine Intelligence}, vol.~3, pp.~1--20, 2021.

\bibitem{ai2022decompositional}
X.~Ai, Z.~Zhang, L.~Sun, J.~Yan, and E.~Hancock, ``Towards quantum graph neural networks: An ego-graph learning approach,'' {\em arXiv preprint arXiv:2201.05158}, 2022.

\bibitem{zhang2019quantum}
Z.~Zhang, D.~Chen, J.~Wang, L.~Bai, and E.~R. Hancock, ``Quantum-based subgraph convolutional neural networks,'' {\em Pattern Recognition}, vol.~88, pp.~38--49, 2019.

\bibitem{zheng2021quantum}
J.~Zheng, Q.~Gao, and Y.~L{\"u}, ``Quantum graph convolutional neural networks,'' in {\em 2021 40th Chinese Control Conference (CCC)}, pp.~6335--6340, IEEE, 2021.

\bibitem{chen2021hybrid}
S.~Y.-C. Chen, T.-C. Wei, C.~Zhang, H.~Yu, and S.~Yoo, ``Hybrid quantum-classical graph convolutional network,'' {\em arXiv preprint arXiv:2101.06189}, 2021.

\bibitem{xu2023quantum}
S.~Xu, F.~Wilhelm-Mauch, and W.~Maass, ``Quantum feature embeddings for graph neural networks,'' in {\em Hawaii International Conference on System Sciences}, 2023.

\bibitem{schlingemann2001stabilizer}
D.~Schlingemann, ``Stabilizer codes can be realized as graph codes,'' {\em arXiv preprint quant-ph/0111080}, 2001.

\bibitem{schlingemann2001quantum}
D.~Schlingemann and R.~F. Werner, ``Quantum error-correcting codes associated with graphs,'' {\em Physical Review A}, vol.~65, no.~1, p.~012308, 2001.

\bibitem{gottesman1997stabilizer}
D.~Gottesman, {\em Stabilizer codes and quantum error correction}.
\newblock California Institute of Technology, 1997.

\bibitem{hein2006entanglement}
M.~Hein, W.~D{\"u}r, J.~Eisert, R.~Raussendorf, M.~Nest, and H.-J. Briegel, ``Entanglement in graph states and its applications,'' {\em arXiv preprint quant-ph/0602096}, 2006.

\bibitem{daskin2018simple}
A.~Daskin, ``A simple quantum neural net with a periodic activation function,'' in {\em 2018 IEEE International Conference on Systems, Man, and Cybernetics (SMC)}, pp.~2887--2891, IEEE, 2018.

\bibitem{defferrard2016convolutional}
M.~Defferrard, X.~Bresson, and P.~Vandergheynst, ``Convolutional neural networks on graphs with fast localized spectral filtering,'' {\em Advances in neural information processing systems}, vol.~29, 2016.

\bibitem{daigavane2021understanding}
A.~Daigavane, B.~Ravindran, and G.~Aggarwal, ``Understanding convolutions on graphs,'' {\em Distill}, vol.~6, no.~9, p.~e32, 2021.

\bibitem{daskin2022quantum}
A.~Daskin, ``Quantum implementation of circulant matrices and its use in quantum string processing,'' {\em arXiv preprint arXiv:2206.09364}, 2022.

\end{thebibliography}
\end{document}